\documentclass[romp]{article}
\usepackage{amsfonts,amssymb,amsmath,amsthm,cite}
\usepackage{graphicx}
\DeclareMathOperator{\Res}{Res}         
\DeclareMathOperator{\Tr}{Tr}                 
\newtheorem{assumption}{Assumption}[section]
\newtheorem{theorem}[assumption]{Theorem}
\newtheorem{corollary}[assumption]{Corollary}

\newtheorem{lemma}[assumption]{Lemma}
\newtheorem{definition}[assumption]{Definition}
\newtheorem{prop}[assumption]{Proposition}
\newtheorem{remark}[assumption]{Remark}

\newcommand{\A}{\mathcal{A}}              
\renewcommand{\a}{\alpha}                    
\newcommand{\B}{\mathcal{B}}              
\newcommand{\C}{\mathbb{C}}              
\newcommand{\DD}{\mathcal{D}}           
\renewcommand{\H}{\mathcal{H}}          
\newcommand{\K}{\mathcal{K}}             
\renewcommand{\L}{\mathcal{L}}          
\newcommand{\<}{\langle}       
     
\newcommand{\N}{\mathbb{N}}            
\newcommand{\OO}{\mathcal{O}}

\newcommand{\Q}{\mathbb{Q}}             
\newcommand{\R}{\mathbb{R}}             

\newcommand{\set}[1]{\{\,#1\,\}}              
\renewcommand{\SS}{\mathcal{S}}        
\newcommand{\T}{\mathbb{T}}                
\newcommand{\U}{\mathcal{U}}              
\newcommand{\Z}{\mathbb{Z}}                 

\newcommand{\bb}{\begin{eqnarray}}
\newcommand{\ee}{\end{eqnarray}}
\newcommand{\eee}{\nonumber\end{eqnarray}}

\def\<#1,#2>{\langle#1\,,\,#2\rangle}      
\newbox\ncintdbox \newbox\ncinttbox
\setbox0=\hbox{$-$} \setbox2=\hbox{$\displaystyle\int$}
\setbox\ncintdbox=\hbox{\rlap{\hbox
    to \wd2{\kern-.1em\box2\relax\hfil}}\box0\kern.1em}
\setbox0=\hbox{$\vcenter{\hrule width 4pt}$}
\setbox2=\hbox{$\textstyle\int$}
\setbox\ncinttbox=\hbox{\rlap{\hbox to \wd2{\kern-.14em\box2\relax\hfil}}\box0\kern.1em}
\newcommand{\ncint}{\mathop{\mathchoice{\copy\ncintdbox}
    {\copy\ncinttbox}{\copy\ncinttbox}
    {\copy\ncinttbox}}\nolimits}
\title{Compact $\kappa$-deformation and spectral triples}
\author{B. Iochum\thanks{Also at Universit\'e de Provence} , 
T. Masson, T. Sch\"ucker$^{*}$, \\
Centre de Physique Th\'eorique\thanks{%
UMR 6207: Unit\'e Mixte de Recherche du CNRS et des Universit\'es Aix-Marseille I, Aix-Marseille II et de l'Universit\'e du Sud Toulon-Var. Laboratoire affili\'e \`a la FRUMAM -- FR 2291} , CNRS--Luminy, \\ 
Case 907, 13288 Marseille Cedex 9, FRANCE \\[2ex]
A. Sitarz\thanks{Partially supported by MNII grant 189/6.PRUE/2007/7 and N 201 1770 33.} \\
Institute of Physics, Jagiellonian University,\\
Reymonta 4, 30-059 Krak\'ow, Poland
\phantom{\footnote{Emails: iochum@cpt.univ-mrs.fr, masson@cpt.univ-mrs.fr, thomas.schucker@gmail.com, sitarz@if.uj.edu.pl}} 
}

\begin{document}
\maketitle
\begin{abstract}
We construct discrete versions of $\kappa$-Minkowski space related to a certain compactness of the time coordinate. We show that these models fit into the framework of noncommutative geometry in the sense of spectral triples. The dynamical system of the underlying discrete groups (which include some Baumslag--Solitar groups) is heavily used in order to construct \emph{finitely summable} spectral triples. This allows to bypass an obstruction to 
finite-summability appearing when using the common regular representation. The dimension of these spectral triples  is unrelated to the number of coordinates defining the $\kappa$-deformed Minkowski spaces.
\end{abstract}




\section{Introduction}

In 1991, Lukierski, Ruegg, Nowicki \& Tolstoi \cite{kappafirst1,kappafirst2} discovered a Hopf algebraic 
deformation of the Poincar\'e Lie algebra. The deformation parameter with units of mass is 
traditionally called $\kappa $. Only two years later, the Hopf algebra was represented on the 
$\kappa $-deformation of Minkowski space \cite{kappaMink1,kappaMink2}. This noncommutative space 
has been used immediately to generalize the notion of a quantum particle \cite{part}. Its use to generalize 
quantum fields is still in progress \cite{fields}. Noncommutative differential calculi on the 
$\kappa $-Minkowski space were investigated \cite{calculi} as well as the Noether theorem 
\cite{noether,{AmelinoCamelia:2007rn}}. The $\kappa $-Minkowski space has been popularized as `double
special relativity' \cite{double} and is also known to appear in a 3-dimensional spinfoam model \cite{foam}.

It is natural to ask whether $\kappa $-Minkowski space is a noncommutative geometry in the sense 
of Alain Connes \cite{triple,triple1}. So far, despite some attempts  \cite{answer}, this question remained open.
\vspace{0.5cm}

In the $\kappa$-deformation of $n$-dimensional Minkowski space, the
space-time coordinates satisfy the following solvable Lie-algebraic relations:
\begin{align}
\label{commutationkappa}
[x^0,\,x^j]:= \tfrac{i}{\kappa} \,x^j, \quad [x^j,\,x^k]=0, \quad j,k=1,\dots,n-1.
\end{align}
Here we assume $\kappa>0$.

Using the Baker--Campell--Hausdorff formula, one gets \cite[eq. (2.6)]{KMLS}
\begin{align*}
e^{i c_\mu x^\mu}=e^{i c_0 x^0}\, e^{i c'_j \,x^j} \text{ where } 
c'_j:= \tfrac{\kappa}{c_0}\,(1-e^{-c_0/\kappa}) c_j.
\end{align*}
Actually, if $[A,B]=sB$, we have the ``braiding identity"
\begin{align}
\label{braiding}
 e^A \, e^B=e^{(\exp s)B}\,e^A.
 \end{align}
If the $x^\mu$'s are selfadjoint operators on some Hilbert space, we can define the $n$ 
unitaries  
$$
U_\omega :=e^{i\omega x^0} \text{ and }V_{\vec{k}}:=e^{-i\sum_{j=1}^{n-1} k_jx^j}
$$
with $\omega,k_j \in \R$, which generate the $\kappa$-Minkowski group considered in 
\cite{Agostini} and an algebra $\A_\kappa$.

 If $W(\vec{k},\omega):=V_{\vec{k}}\,U_\omega$, one gets as in \cite[eq. (13)]{Agostini}
\begin{align}
\label{grouplaw}
 W(\vec{k},\omega) \, W( \vec{k'},\omega')=W(e^{-\omega/\kappa} \vec{k'}+\vec{k},\omega + \omega').
\end{align}

The group law \eqref{grouplaw} is, for $n=2$, nothing else but the crossed product 
\begin{align}
\label{groupkappa}
G_\kappa:=\R \rtimes_\a \R \text{ with group isomorphism }\,\a(\omega)k:=e^{-\omega/\kappa}k, \,\,\, k\in \R.
\end{align}
$G_\kappa \simeq \R\rtimes \R^*_+$ is the affine group on the real line which is solvable and nonunimodular 
(left Haar measure is $e^{\omega/\kappa}d\omega\,dk$). Being connected and simply connected, it is 
uniquely determined by its Lie algebra \eqref{commutationkappa}. All its irreducible unitary 
representations are known: they are either one-dimensional, or fall into two inequivalent classes 
\cite{Agostini}.

It is worthwhile to recall that, when $\kappa \rightarrow \infty$, the usual plane $\R^2$ is recovered but 
with an unpleasant pathology at the origin \cite{DP}.

An interesting situation occurs  when $m :=e^{-\omega /\kappa}  \in \N^*$ for a given $\omega$, since the group 
law can be reduced to a commutation relation between unitaries:
\begin{align}
\label{commutationgen}
 U_\omega V_{\vec{k}}=(V_{\vec{k}})^m U_\omega,
 \end{align}
 $m$ being independent of $k_j$.

\bigskip

In this work, we want to investigate different spectral triples ($\A,\H,\DD)$ associated to the group 
$C^*$-algebra of $G_\kappa$.  To avoid technicalities due to a continuous spectrum of $\DD$, we want a 
unital algebra so that we consider a periodic time $x^0$ which induces a discrete version $G_a$ of $G_\kappa$ 
where $a$ is a real parameter depending on $\kappa$. This is motivated and done in section \ref{motivation}.
 
In section \ref{aquelconque}, we give the main properties of the algebra $\U_a = C^*(G_a)$ and its representations. 
For $C^*$-algebras of groups, the left regular representation is the natural one to consider. But due to the structure 
of $G_a$ (solvable with exponential growth, see Theorem~\ref{Thealgebra}), there is a well-known obstruction to 
construct finite-summable spectral triples on $\U_a$ based on this representation. In order to bypass this obstruction,  
we need to refine our understanding of the structure of $G_a$ in terms of an underlying dynamical system. This 
structure permits to define in \ref{representationsinduced} a particular representation of $\U_a$ using only the 
periodic points of the dynamical system. At the same time, following Brenken and J\o rgensen \cite{BJ}, the 
topological entropy of this dynamical system is considered. 

It is worthwhile to notice that the elementary building blocks $G_a$ 
given by $a=m \in \N^*$ as in \eqref{commutationgen} are just some of the amenable  Baumslag--Solitar groups, 
already encountered in wavelet theory \cite{J}. The power of harmonic analysis on groups also justifies a reminder 
of their main properties in section \ref{abstractapproach}. 

Finally, different spectral triples are exhibited in section \ref{exist}. The question of the finite summability of these 
triples is carefully considered with results by Connes \cite{Con89} and Voiculescu \cite{Voiculescu79, Voiculescu}: 
using the regular representation of $C^*(G_a)$, an obstruction to finite-summability appears and allows only 
$\theta$-summability.
However, representations of $C^*(G_a)$, not quasi-equivalent to the left regular one and based on the 
existence of periodic points for dynamical systems, can give rise to arbitrary finite-summable spectral triples. These 
results are summarized in Theorems \ref{nonexistence} and \ref{existence} and they are commented in subsequent 
remarks.

\section{Motivations and models}
\label{motivation}

Relations as in \eqref{commutationkappa} define an abstract noncommutative algebra in the generators 
$x^0, x^j$. As a purely algebraic object, this algebra contains elements which are finite sums of 
monomials in the generators. However, this construction is not suitable to study the noncommutative geometry 
(in the approach of Connes) of the $\kappa$-deformed (noncommutative) Minkowski space.

A much better approach is to consider a topological $C^*$-algebra, which will be based on the relations 
\eqref{commutationkappa}, and its representations on a Hilbert space. Certainly, then the operators  $x^0, x^j$ 
can be expected to be unbounded. The usual procedure is to use the Weyl quantization. 
In the following, we give a brief review of the method illustrated by three examples. This will help to understand 
the algebra we will consider later as our model in section \ref{themodel-subgroup}. For simplicity, we restrict 
ourselves to the two-dimensional case.

The first example is the undeformed plane $\R^2$. In the language of noncommutative geometry, one can 
(naively) describe its algebra as generated by two commuting hermitian generators $\hat{x}$ and $\hat{y}$. 
We suppose that these generators are 
represented as (unbounded) operators on a Hilbert space $\H$. To reduce cumbersome notations, we will  
use the same symbols $\hat{x}$ and $\hat{y}$ for the operators so that the following manipulations take 
place in the algebra of operators on $\H$. For any $(a,b) \in \R^2$, we introduce the unitary operators 
$W(a,b) := e^{i a \hat{x}} e^{i b \hat{y}}$. To any function $f \in L^1(\R^2)$, one associates the bounded 
operator on $\H$
\begin{equation*}
W(f) := \int_{\R^2} f(a,b) \,W(a,b) \, da \,db\,.
\end{equation*}
This expression looks very much like a Fourier transform of $f$ to which it is obviously related. The algebra 
we are looking for, is then a topological algebra generated by the $W(f)$'s for $f$ in a convenient space of 
functions: to be precise, we consider the $C^*$-algebra generated by the $W(f)$'s for $f\in L^1(\R^2)$.

This construction yields a product $f \ast g$ of two functions $f, g \in L^1(\R^2)$, which is defined by the 
relation $W(f \ast g) := W(f) W(g)$ and in order to understand this product, let us digress and 
show how this construction can be related to well known facts about group $C^*$-algebras. 

Let $G$ be a topological group and let $W$ be a unitary representation of $G$ on a Hilbert space $\H$. 
Then, one has $W(s)W(t) = W(st)$ as unitary operators for any $s, t \in G$. To  
$f \in L^1(G, d\mu)$ (where $d\mu$ is a left Haar measure on $G$), one associates  
$W(f) := \int_G f(s)\, W(s) \, d\mu(s)$ which is a bounded operator on $\H$. 
Then, the $C^*$-algebra generated by the $W(f)$ for 
$f \in L^1(G, d\mu)$ is the image of the representation induced by $W$ of the reduced $C^*$-algebra 
$C^*_{red}(G)$ of $G$ on $\H$. Indeed,
\begin{equation*}
W(f) \, W(g) = \int_G \left( \int_G f(t) g(t^{-1}s)  \, d\mu(t) \right) W(s) \, d\mu(s),
\end{equation*}
so, if $(f \ast g)(s) := \int_G f(t) g(t^{-1}s)  \, d\mu(t)$ is the usual convolution product in $C^*_{red}(G)$, then $W$ 
is an algebra homomorphism. 

Let us return to the example of the plane. One has $W(a,b)W(a',b') = W(a+a', b+b')$ so that $W$ is a 
representation of the abelian group $G = \R^2$ for the additive law on both factors. This implies that the 
$C^*$-algebra we are looking for in that case is a representation of $C^*(\R^2) \simeq C_0(\R^2)$, where 
the explicit isomorphism is obtained by Fourier transforms. Here $C^*(\R^2)$ is an algebra of functions 
in the variables $(a,b) \in \R^2$ for the (ordinary) convolution product on $\R^2$ while $C_0(\R^2)$ is an 
algebra of functions (continuous and vanishing at infinity) in the variables $(x,y)$ (the spectrum of the 
operators representing $\hat{x}$ and $\hat{y}$) for the ordinary product of functions. Notice that as expected
 the generators $\hat{x}$ and $\hat{y}$ do not belong to this algebra.

\medskip
The second example is the space with canonical commutation relations. This noncommutative space is 
generated by two hermitian elements $\hat{p}$ and $\hat{q}$ such that $[\hat{q}, \hat{p}] = i \hbar$. 
Then the usual Weyl procedure defines the unitary operators 
$W(\alpha, \beta) := e^{i \alpha \hat{q} + i \beta \hat{p}}$ 
for any $(\alpha, \beta) \in \R^2$. This gives immediately
\begin{equation}
\label{eq-weylCCR}
W(\alpha, \beta) \, W(\alpha', \beta') = e^{-\frac{i \hbar}{2}(\alpha \beta' - \beta \alpha')} \,W(\alpha + \alpha', 
\beta + \beta').
\end{equation}
The associated $C^*$-algebra is generated by the bounded operators 
\begin{equation*}
W(f) := \int_{\R^2} f(\alpha, \beta) \, W(\alpha, \beta) \, d\alpha \, d\beta
\end{equation*}
for $f \in L^1(\R^2, d\alpha\, d\beta)$. Because \eqref{eq-weylCCR} is not a group representation, this 
$C^*$-algebra is not related to the group $C^*$-algebra. In fact, \eqref{eq-weylCCR} expresses the fact 
that this is a projective representation of the abelian group $\R^2$. One can transform it into a 
representation of the Heisenberg group $H_3$ which is the convenient central extension of $\R^2$ to do 
this end. Then the $C^*$-algebra associated to the canonical commutation relations is related to the 
$C^*$-algebra of the Heisenberg group $H_3$.

\medskip
Let us now consider the example given by \eqref{commutationkappa} for two hermitian generators $x^0$ 
and $x^1$. For any $(k,\omega) \in \R^2$, one defines as before 
$W(k,\omega) := V_{k} \,U_\omega = e^{-ik x^1} e^{i \omega x^0}$. Then one has \eqref{grouplaw} which is
a representation of the non-abelian group $G_\kappa$ defined in \eqref{groupkappa}. The $C^*$-algebra 
generated by the bounded operators
\begin{equation*}
W(f) := \int_{G_\kappa} f(k, \omega) \,W(k, \omega) \, e^{\omega/\kappa}  \,dk \,d\omega
\end{equation*}
for any $f \in L^1(G_\kappa, e^{\omega/\kappa}  \, dk\,d\omega)$ (here $e^{\omega/\kappa} dk\,d\omega$ is 
the left Haar measure on $G_\kappa$) is the representation of $C^*_{red}(G_\kappa)$ by $W$. 
The explicit product of two elements $f, g \in L^1(G_\kappa, e^{\omega/\kappa}  \, dk\,d\omega)$ takes the 
form
\begin{equation*}
(f \ast_\kappa g)(k, \omega) = \int_{G_\kappa} f(k', \omega') g(e^{\omega'/\kappa}(k - k'), \omega - \omega') 
\,e^{\omega'/\kappa} \,dk'\,d\omega'.
\end{equation*}
As for the example of the plane, the generators $x^0$ and $x^1$ cannot belong to this algebra.

The topological algebra to consider in order to study the noncommutative geometry of the $2$-dimensional space 
defined by \eqref{commutationkappa}, is then the $C^*$-algebra $C^*_{red}(G_\kappa)$ or some of its 
dense subalgebras.

\medskip
The advantage of considering the theory of group $C^*$-algebras is twofold. Many structural 
properties on groups will turn out to be useful in studying some properties of the corresponding $C^*$-algebras. 
This is why we will focus on groups in the following. Furthermore, this allows us to construct in a natural way 
compact versions of noncommutative spaces as we shall now explain.

For an abelian topological group $G$, $C^*_{red}(G)$ is isomorphic to $C_0(\widehat{G})$ where 
$\widehat{G}$ is the Pontryagin dual of $G$. Both algebras are defined as spaces of functions. By duality, a 
discrete subgroup $\Gamma \subset G$ produces the $C^*$-algebra 
$C^*_{red}(\Gamma) \simeq C(\widehat{\Gamma})$ where $C(\widehat{\Gamma})$ is the $C^*$-algebra of 
continuous functions on the compact space $\widehat{\Gamma}$. Notice that there is a natural dual map 
$\widehat{G} \rightarrow \widehat{\Gamma}$. For the example of the plane, consider the discrete subgroup 
$\Gamma = \Z^2 \subset \R^2$. Then the resulting $C^*$-algebra is $C^*(\Z^2) \simeq C(\T^2)$ because 
$\widehat{\Z} = \T^1$. The dual map $\R \simeq \widehat{\R} \rightarrow \widehat{\Z} = \T^1$ is explicitly 
given by $x \mapsto e^{2\pi ix}$. The choice of the subgroup $\Gamma = \Z^2 \subset \R^2$ corresponds 
then to the choice of the compact version $\T^2$ of the (dual) space $\widehat{\R}^2 \simeq \R^2$. The 
compactification takes place in the space of the variables $(x,y)$. Notice that not all compactifications of $\R^2$ 
can be obtained this way. The $2$-sphere is a counter-example.

This procedure cannot be applied directly to the canonical commutation relations because its  
$C^*$-algebra is not a group $C^*$-algebra. But, as we will show, this procedure can be applied with 
success to the $\kappa$-deformed Minkowski space. In order to get a compact version of this noncommutative 
space, one has to choose a discrete subgroup $H_\kappa \subset G_\kappa$. Since $H_\kappa$ is discrete and 
non-abelian, the associated algebra $C^*(H_\kappa)$ is unital and noncommutative, so it can be interpreted as 
a compact noncommutative space. This point is motivated in section \ref{spectraltriples} and is done in section 
\ref{themodel-subgroup}.

As a final preliminary remark, let us mention that the groups we will encounter will be decomposed as crossed 
products with $\Z$, so both the (related) theories of discrete dynamical systems and crossed products of 
$C^*$-algebras will be intensively used in many parts of this work.

\subsection{Spectral triples}
\label{spectraltriples}
The goal of this work is to study the existence of spectral triples for the $\kappa$-deformed space. Recall that a 
spectral triple (or unbounded Fredholm module) $(\A,\H,\DD)$ \cite{Con95,triple,ConnesMarcolli} is given by a 
unital $C^*$-algebra $A$ with a faithful representation $\pi$ on a Hilbert space $\H$ and an unbounded
self-adjoint operator $\DD$ on $\H$ such that 

\qquad - the set $\A=\set{a \in \A \,: \, [\DD, \pi(a)] \text{ is bounded }}$ is norm dense in $A$,

\qquad - $(1+\DD^2)^{-1}$ has a compact resolvent.

($\A$ is always a $^*$-subalgebra of $A$.)

Of course, the natural choice of the algebra $A$ is to take the $C^*$-algebra of the group $G_\kappa$, but since
$A=C^*(G_\kappa)$ has no unit, we need to replace the second axiom by:

\qquad - $\pi(a) (1+\DD^2)^{-1}$ has a compact resolvent for any $a\in \A$.

This technical new axiom generates 
a lot of analytical complexities but is necessary to capture the metric dimension associated to $\DD$. For 
instance, if a Riemannian spin manifold $M$ is non-compact, the usual Dirac operator $\DD$ has a 
continuous spectrum on $\H=L^2(S)$ where $S$ is the spinor bundle on $M$. Nevertheless, the spectral  
triple $\big(C^\infty(M),  L^2(S),\DD\big)$ has a metric dimension which is equal to the dimension of $M$.
A noncommutative example (the Moyal plane) of that kind has been studied in \cite{GIGSV}.

We try to avoid these difficulties here using a unital algebra $A$.

\subsection{The compact version model as choice of a discrete subgroup}
\label{themodel-subgroup}

In this paper, we consider only dimension $n=2$, since many of the results can be extended to higher 
dimensions thanks to 
\eqref{commutationgen}.

As explained in details above, in order to get a unital $C^*$-algebra, we prefer to work with a discrete 
subgroup $H_\kappa$ of $G_\kappa$ yielding such that $1\in C^*(H_\kappa)$.

Since we want also to keep separate the role of the two variables $x^0$ and $x^1$, it is natural to consider 
the subgroup of the form  $H_\kappa =H \rtimes_\a \Z$: we first replace the second $\R$ of $G_\kappa$ in 
\eqref{groupkappa} by the lattice $\Z$ which corresponds to unitary periodic functions of a chosen frequency 
$\omega_0$ (the time $x^0$ is now periodic). 

So, given $\kappa>0$ and $\omega_0 \in \R$, with 
\begin{align}
\label{defa}
a:=e^{-\omega_0 /\kappa} \in \R^+,
\end{align}
the group $\R\rtimes_{\a_a} \Z$ is a subgroup of $G_\kappa$ where $\a_a(n)$ is the multiplication by 
$a^n$. This subgroup is the affine group Aff$_1(A)$ of a commutative unital ring $A=\R$ (or ``$ax+b$'' group). 
It is not discrete. 

Then we want a group $H$ to be a discrete (now, not necessarily topological) subgroup of the first $\R$ in 
$\R\rtimes_{\a_a} \Z$, which is invariant by the action $\a_a$. Given $k_0 \in \R$, a natural building block 
candidate for a discrete $H$ is given by 
$H = B_a \cdot k_0 \simeq B_a$ where 
$$
B_a:=\set{\sum_{i, \,{\rm finite}} m_i\, a^{n_i}  \, : \, m_i,n_i \in \Z}.
$$
More generally, one can take $H\simeq \oplus_{k_0\in S} \,B_a$ where $S$ is a discrete 
subspace of $\R^+$ such that $B_a \cdot k_0 \cap B_a \cdot k_1= \varnothing$ for any $k_0,k_1 \in S$.

In some way, the search for a discrete subgroup $H_\kappa$ of $G_\kappa$ such that $1 \in C^*(H_\kappa)$ 
leads us to the group $H_{\kappa, a} :=B_a \rtimes_{\a_a} \Z$ which is isomorphic to a subgroup of $G_\kappa$ 
once $k_0$ is fixed.

This procedure drives us to the following situation which depends in particular upon the algebraic nature of $a$. 
Even if the details are given in section \ref{aquelconque}, it is interesting to quote here that for $a$ 
transcendental, $B_a=\oplus_{\Z} \Z$, so that the groups $H_{\kappa, a}$ are all isomorphic. When $a$ is 
algebraic, $\widehat B_a$ is a solenoid, isomorphic to a subgroup of $(\T^d)^\Z$ where $d$ is related to the 
algebraic nature of the real number $a$. In that case, the structure of the groups $H_{\kappa, a}$ relies heavily 
on $a$.

Finally, simplifications occur when $a=m \in \N^*$ is an integer (which requires $\omega_0<0$, but see the 
remark after Lemma~\ref{symmetry}): this group $H_{\kappa, m}$ is well 
known since it is the solvable Baumslag--Solitar group $BS(1,m)=\Z[\tfrac{1}{m}] \rtimes_{\a_m} \Z$ as shown in 
section \ref{abstractapproach} where we get $B_m=B_{1/m}=\Z[1/m]$. Moreover, in that situation we recover the 
two unitaries $U_{\omega_0}$ and $V_{k_0}$ satisfying \eqref{commutationgen} and we may assume 
$H_{\kappa, m}= B_m \cdot k_0 \rtimes \Z$.

\medskip

To conclude, we see that for different values of the (dimensionless) parameter $a$, we obtain a broad range of 
noncommutative spaces. In some cases, the algebras simplify significantly, as summarized in the following 
lemma:

\begin{lemma}
\label{caseinteger}
In two dimensions, there exists a unital subalgebra $C^*(H_{\kappa, m})$ of the $\kappa$-defor\-ma\-tion 
algebra which is associated to the subgroup $H_{\kappa, m}=\Z[\tfrac{1}{m}] \rtimes_{\a_m} \Z$ of $G_{\kappa, m}$ 
and can be described as the algebra generated by two unitaries $U,V$ satisfying the constraints 
$U=U_{{\omega}_0}$, $V=V_{k_0}$ and $UV=V^mU$. Here,  $\kappa :=-\omega_0\,\log^{-1}(m)>0$ for some 
given integer $m> 1$ and some $\omega_0 \in \R^-$, $k_0\in \R$.
\end{lemma}

\section{The algebra $\U_a$ and its representations}\label{aquelconque}

In this section we define precisely and study in details the $C^*$-algebra $\U_a$ which is our model for a compact 
version of the $2$-dimensional $\kappa$-Minkowski space. The structure of this algebra is described through a 
semi-direct product of two abelian groups, one of which depends explicitly on the real parameter $a >0$. This 
semi-direct structure gives rise to a dynamical system which is heavily used in the following. The classification of the 
algebras $\U_a$ is performed: the $K$-groups are not complete invariants, and we use the entropy defined on the 
underlying dynamical system to complete this classification. Then some representations of $\U_a$ are considered. 
They depends on the algebraic or transcendental character of $a$. In the algebraic case, some particular finite 
dimensional representations are introduced based on periodic points of the dynamical system. This construction will 
be used in section~\ref{exist}.

Let $a=e^{-\omega_0/\kappa} \in \R^*_+$ with $a \neq 1$, and let us recall general facts from \cite{BJ}:

Define
$$
B_a:=\set{ \sum_i    m_i \,a^{n_i}  \text{ for finitely many } m_i,n_i \in \Z  }.
$$
This discrete group is torsion-free so its Pontryagin dual $\widehat B_a$ is connected and compact.  

Let $\a_a$ be the action of $\Z$ on $b \in B_a$ defined by $\a_a(n)b:=a^n \,b$, let $\widehat{\a_a}$ be the 
associate automorphism on $\widehat B_a$ and
\begin{align*}
G_a & :=B_a \rtimes_{\a_a} \Z,
&
\U_a & := C^*(G_a) = C^*(B_a) \rtimes_{\a_a} \Z = C(\widehat B_a) \rtimes_{\widehat{\a_a}} \Z.
\end{align*}
This type of $C^*$-algebras also appeared in \cite{CPPR} for totally different purposes.

We will write $\a$ instead of $\a_a$ when there is no confusion. The group $G_a$ is generated by the two 
generators 
$$
u:=(0,1) \text{ and } v:=(1,0).
$$

\begin{lemma}
\label{symmetry}
Let $a\in \R_+^*$, then $B_a=B_{1/a}$ and 
$G_a \simeq G_{1/a}$. Thus the $C^*$-algebra $\U_a$ and $\U_{1/a}$ are isomorphic.
\end{lemma}

\begin{proof}
Clearly $B_a=B_{1/a}$ as subset of $\R$. The map 
$:(b,n)\in B_a \rtimes_{\a_a} \Z \mapsto (b,-n)\in B_{1/a} \rtimes_{\a_{1/a}} \Z$ 
defines a group isomorphism between $G_a$ and $G_{1/a}$.
\end{proof}

As a consequence, the restriction on the sign of the frequency $\omega_0$ in Lemma \ref{caseinteger} is just an 
artifact since the case $a=m$ and $a=1/m$ are the same.

The symmetry point $a=1/a$ corresponds to the commutative case in \eqref{commutationgen}  
with $a=1$ or the undeformed relation \eqref{commutationkappa} with $\kappa=\infty$. In this spirit $\U_a$ can 
be viewed as a deformation of the two-torus.

\bigskip 
To analyze $\U_a$, we essentially follow \cite{BJ,Brenken}. The dynamical system 
$\U_a\simeq C(\widehat B_a) \rtimes_\a \Z$ has an ergodic action (if $a\neq1$) and a lot of periodic points. If 
$$
\mathrm{Per}_q(\widehat B_a) := \set{\chi  \in \widehat B_a \, : \,  \widehat{\a}^{k}(\chi) \neq  \chi, 
\ \forall  k<q, \ \widehat{\a}^{q}( \chi) =  \chi}
$$
is the set of $q$-periodic points, then the growth rate 
$\lim_{q\rightarrow \infty} q^{-1}\log \big(\# \mathrm{Per}_q(\widehat B_a)\big)$ of this sets is an invariant of $\U_a$ 
which coincides with the topological entropy $h(\widehat{\a_a})$ by Yuzvinskii's formula (the topological entropy 
$h(\gamma)$ is a good invariant for a homeomorphism $\gamma$ of a compact topological space onto itself):
\begin{align}
\label{entrop}
h(\widehat{\a_a})=\lim_{q\rightarrow \infty} q^{-1}\log \big(\# \mathrm{Per}_q(\widehat B_a)\big).
\end{align}
This entropy can be finite or infinite, dividing the algebraic properties of $a$ into two cases: $a$ can be a 
an algebraic or a transcendental number.

\subsection{Transcendental case}
\label{transcendent}

If $a$ is transcendental, then $B_a\simeq \Z[a,a^{-1}]$. Thus, $B_a\simeq \oplus_\Z \Z$, 
$\widehat B_a\simeq S_a:= \set{z=(z_k)_{k=-\infty}^\infty  \in \T^\Z}$ and $\big(\widehat{\a} (z)\big)_k=z_{k+1}$ 
for $z\in S_a, \,k\in \Z$ so that $\widehat{\a}$ is just the shift $\sigma$ on $\T^\Z$. Thus 
$$
\U_a \simeq C(\T^\Z)\rtimes_\sigma \Z \, \text{ and } \,h(\widehat{\a})=\infty.
$$
Note that the wreath product $\wr$ appears with its known presentation:
$$
G_a=B_a \rtimes_\a \Z  \simeq \Z \wr \Z \simeq  \langle u,v \, : \, [u^i v u^{-i},v]=1 \text{ for all } i\geq1 \rangle.
$$
This group is amenable (solvable), torsion-free, finitely generated (but not finitely presented), 
residually finite with exponential growth. 

$S_a$ contains a lot of periodic points: $q$-periodic points are simply obtained by repeating any finite 
sequence $(z_k)_{k=0}^{q-1}$ of arbitrary elements in $\T$. Aperiodic points are easily constructed also.

Moreover, $S_a$ is the Bohr 
compactification $b_{B_a}\R$ of $\R$: let $\hat \iota$ be the dual map of the natural group embedding $\iota$ 
of $B_a$ into $\R$, so $\hat \iota$ is a continuous group homomorphism from $ \R$ to $S_a$ with 
dense image since we have $\hat \iota:x\in \R \mapsto ( z_{k})_{k=-\infty}^\infty \in S_a$ 
with  $z_k= e^{i2\pi a^{-k}x}$. Thus $\hat \iota$ is injective and surjective on elements 
$(z_k)_{k=-\infty}^\infty \in S_a$ such that $\lim_{k\rightarrow \infty}z_k=1$ and it satisfies 
$\hat \iota(ax)=\widehat{\a} \big(\iota(x)\big)$. $\hat \iota$ is an infinitely winding curve on the solenoid $S_a$.

\subsection{Algebraic case}
\label{algebrique}

Assume now that $a$ is algebraic. Let $P\in \Q[x]$ be the monic irreducible polynomial such that $P(a)=0$ and 
let $P=c\,Q_a$ where $c^{-1}$ is the least common multiple of the denominators of  coefficients of $P$ 
(also called content), so $Q_a\in \Z[x]$. Since the coefficients of $Q_a$ have greatest common divisor 1, the 
ideal $(Q_a)$ generated by $Q_a$ in $\Z[x]$ is prime. If $d$ is the degree of $Q_a$, note the following ring 
isomorphism \cite{BJ}
$$
B_a  \simeq \Z[x,x^{-1}]/(Q_a).
$$
Moreover, $B_a$ has a torsion-free rank $d$ since $\set{1,a,a^2,\cdots,a^{d-1} }$ is a 
maximal set of independent elements in $B_a$, and $\widehat B_a$ has dimension $d$.

If 
$$
Q_a(x)=\sum_{j=0}^d q_j\, x^j
$$
(so $Q_a$ has leading coefficient $q_d \in \N^*$), let $A_a \in M_{d\times d}(\Z)$ 
be the $d\times d$-matrix defined by $({A_a})_{i,j}:=q_d \,\delta_{i,j-1}$ for $1\leq j\leq d $ and 
$(A_a)_{d,j}=-q_{j-1}$. Then $q_d\,a^j=\sum_{k=1}^d (A_a)_{j,k}\, a^{k-1}$.

For instance, if $a=1/m$ for $m\in \N^*$ then $P(x)=x-1/m$, $Q_a(x)=mx-1$, $d=1$, so $B_a=\Z[\frac{1}{m}]$ 
and $A_a$ is just the number $1$.

Let $\sigma$ be the shift on the group ${(\T^d)}^\Z$ and consider its $\sigma$-invariant subgroup 
$$
K_a:=\set{z=(z_k)_{k=-\infty}^\infty \in {(\T^d)}^\Z \, : \, q_d\,z_{k+1}=A_a \,z_k}
$$
where we use the identification $\T \simeq \R/\Z$. This group $K_a$ is connected if and only if the greatest 
common divisor of $q_d$ and $\det(A)$ is 1 by\cite[Proposition 3.12]{Brenken}. This is automatic for $d=1$ 
and for $d\geq 2$, this means $q_d$=1 since $\det(A_a)=(-1)^d q_0 q_d^{d-1}$.

If $S_a$ is the connected component of the identity of $K_a$, then there 
exists a topological group isomorphism $\psi \, : \, \widehat B_a \rightarrow S_a$ such that 
$\sigma_{\vert {S_a}} \circ \psi = \psi \circ \widehat{\a}$ \cite[Theorem 19]{Lawton}. For any 
$\chi \in \widehat B_a$, the associated $z=(z_k)_{k=-\infty}^\infty$ is explicitly given by
\begin{equation*}
z_k^{(i)} = \chi(a^{k+i-1})
\end{equation*}
where $z_k = (z_k^{(i)})_{i=1}^{d} \in \T^d$. In particular $z_0$ is given by 
$\big(\chi(1), \chi(a), \dots, \chi(a^{d-1})\big)$. This map is only surjective on the connected 
component of the identity.

When $a=1/m$ with $m\in \N^*$, we will recover $S_{1/m}=S_m$ in \eqref{solenoid}.

There is a natural morphism of groups $\hat{\iota} : \R^d \rightarrow S_a$ defined as follows: 

to any $\phi = (\phi^{(i)})_{i=1,\dots,d} \in \R^d$, one associates 
$\hat{\iota}(\phi) = z = (z_k)_{k=-\infty}^\infty \in {(\T^d)}^\Z$ with
\begin{equation}
\label{eq-thetamap}
z_k^{(i)} = \exp \Big( 2 i \pi q_d^{-k} \sum_{j=1}^{d} (A_a^k)_{i,j} \phi^{(j)} \Big).
\end{equation}
This morphism shows that $S_a$ is a Bohr compactification of $\R^d$ \cite[Proposition 2.4]{Brenken}.

Then $\widetilde{\a}(\phi) = q_d^{-1} A_a \phi$ defines a natural action of $\Z$ on $\R^d$ which 
satisfies $\hat{\iota} \circ \widetilde{\a} = \widehat{\a} \circ \hat{\iota}$.

If $r_i$, $i=1,\cdots d$, are the complex roots of $P$, then by \cite[Proposition 3 and Corollary 1]{BJ},
\begin{align}
\label{cq}
c_q(a):=\#\text{ Per}_q(S_a)=\Pi_{k=1}^q \vert Q_a(e^{i2\pi k/q})\vert 
= \vert q_d \vert^q \, \Pi_{k=1}^d \vert 1-{r_k}^q \vert.
\end{align} 
Thus by \eqref{entrop}, the topological entropy is
\begin{align}
\label{entropy}
h(\widehat{\a_a})=\log\vert q_d \vert +\sum_{i, \,\vert r_i\vert >1} \log \vert r_i\vert.
\end{align}
In case $a=m$ or $a=1/m$, \eqref{entropy} gives $h(\widehat{\a_m})=h(\widehat{\a_{1/m}})=\log(m)$. 

Remark that $c_q(a)  \neq 0$ or $\widehat{\a_a}$ is expansive or $h(\widehat{\a_a}) \neq 0$ if and only if 
$a \neq 1$ or $\vert r_k \vert \neq 1$ for all $k$.

Aperiodic points in $S_a$ can be easily constructed using the map $\hat{\iota}$ defined by \eqref{eq-thetamap}: 
any $\phi \in (\R\backslash \Q)^d$ defines an aperiodic point $\hat{\iota}(\phi) \in S_a$.

\subsection{The structure of algebras $\U_a$}
In order to better understand the structure of the algebra $\U_a$, we obtain some results about the 
underlying dynamical system.

\begin{prop}
\label{prop-dynsyst}
Let $a \in \R^*_+$ and $a\neq 1$. 

(i) The subgroup of periodic points and the set of aperiodic points of $S_a$ under $\widehat{\a}$ are dense. 

(ii) The space of orbits of $S_a$ is not a $T_0$-space.
\end{prop}

\begin{proof}
$(i)$ By \cite[Proposition 4]{BJ}, the periodic points are dense. 

When $a$ is transcendental, the aperiodic points are clearly dense.

For $a \neq 1$ algebraic, we use the fact that the set of aperiodic points is dense if and only if Per$_q(S_a)$ 
has empty interior for any positive integers $q$ by \cite[Lemma 3.1]{ST}. When $a$ is an algebraic number, we 
already saw in \eqref{cq} that Per$_q(S_a)$ has only a finite number of points $c_q(a)$, so has empty interior.

$(ii)$ It is enough to find two points $z$ and $z'$ with different orbits $\mathcal{O}(z) \neq \mathcal{O}(z')$ such 
that $\overline{\mathcal{O}(z)}= \overline{\mathcal{O}(z')}=S_a$, \textsl{i.e.} to find two distinct points in the 
space of orbits whose closures are the same. We will use the following criteria: a subset 
$X \subset \widehat B_a$ is dense in $\widehat B_a$ if and only if 
$\set{b \in B_a : \chi(b) = 1, \chi \in X†} = \set{ 0 }$.

Consider first the case $a \neq 1$ algebraic with associated irreducible polynomial $Q_a \in \Z[x]$. Let 
$\phi \in \R^d$ and define $z = \hat{\iota}(\phi)$. Using the relations induced by $Q_a$ on the monomials $a^r$ for 
$r \in \Z$, any element $b = \sum_{i \in I} m_i a^i$ with $I \subset \Z$ finite and $m_i \in \Z$, can be written 
$b = \sum_{j=1}^{d} r_j a^{j-1}$ with now $r_j \in \Q$. Then 
$z(b) = \exp\left( 2i \pi q_d^{-k} \sum_{j=1}^{d} r_j \phi^{(j)}\right)$. For the particular points $z = \hat{\iota}(\phi)$ with 
$\phi^{(j)} = \phi^{(0)} \in \R \backslash \Q$, the relation $z(b)=1$ admits no non trivial solution in $b$, so that 
$b=0$.

When $a$ is transcendental, the situation is simpler. Consider the aperiodic point 
$z$ defined by $z_k = e^{2 i \pi k\alpha}$ for an irrational $\alpha \in \R \backslash \Q$. Then one has 
$z(b) = \exp \left( 2i \pi \alpha \sum_{k \in I} k \,m_k \right)$ and only $b =0$ is solution to $z(b)=1$.

In both cases, the orbits of all given points $z$ are dense in $S_a$, and there are a lot of such points with 
different orbits, so the space of orbits of $S_a$ is not a $T_0$-space.
\end{proof}

In this proof we saw that many points $z \in S_a$ have dense orbits in $S_a$.

We can now give the main properties of the algebras $\U_a$:
\begin{theorem}
\label{Thealgebra}
Let $a \in \R^*_+$ and $a\neq 1$. Then

(i) The group $G_a = B_a \rtimes _\a \Z$ is a torsion-free discrete solvable group with exponential growth and 
$\widehat{B_a}$ is a compact set isomorphic to a solenoid $S_a$.

(ii) $\U_a=C^*_{red}(B_a \rtimes_\a \Z)\simeq C(S_a) \rtimes_{\widehat{\a}} \Z$ is a NGCR
\footnote{A $C^*$-algebra $A$ is said to be CCR or liminal if $\pi(A)$ is equal to the set of compact operators on 
the Hilbert space $\H_\pi$ for every irreducible representation $\pi$. The algebra $A$ is called NGCR if it has no 
nonzero CCR ideals. \\
An $AF$-algebra is a inductive limit of sequences of finite-dimensional $C^*$-algebras. \\
The algebra $A$ is residually finite-dimensional if it has a separating family of finite-dimensional representations.}, 
AF-embeddable, non-simple, residually finite dimensional $C^*$-algebra and its generated von Neumann 
algebra for the left regular representation is a type $\mathrm{II}_1$- factor. 

(iii) Every element in $\U_a$ has a connected spectrum. In particular, the only idempotents in 
$\U_a$ are 0 and 1.

(iv) The $C^*$-subalgebra $C(S_a)$ is maximal abelian in $\U_a$.
\end{theorem}

\begin{proof}
$(i)$ It remains to control the growth of the group. We apply a method inspired by the case of Baumslag--Solitar 
groups, using the fact that $G_a$ is generated by the two elements $u=(0,1)$ and $v=(1,0)$. Define the map 
$\pi:\, G_a \rightarrow GL(2,\R)$ with
\begin{equation}
\label{eq-repGaGL2R}
\pi(u):=  \left( \begin{array}{cc}
a &  0\\
0 & 1
\end{array} \right), \qquad
\pi(v):=  \left( \begin{array}{cc}
1 & 1\\
0  & 1\\
\end{array} \right).
\end{equation}
Then, $\pi$ is a morphism of groups and $\pi(G_a)$ has an exponential growth as in 
\cite[Example 3, p. 188]{Harpe}. 
Since the growth of a quotient of a group is less than the growth of the group itself, we get the result.

$(ii)$ To show that $\U_a$ is NGCR, it is sufficient, using \cite[Theorem 8.43]{Williams}, to show that the orbit 
space of $S_a$ is not a $T_0$-space. This is done in $(ii)$ of Proposition~\ref{prop-dynsyst}.

The set of periodic points $z$ in $S_a$ is dense by Proposition~\ref{prop-dynsyst} and the non-wandering 
set of $(S_a,\widehat{\a})$ is $S_a$ \cite{BJ}. Thus $\U_a=C(S_a) \rtimes_{\widehat{\a}} \Z$ is $AF$-embeddable 
by \cite{Pimsner} and so, $\U_a$ is quasidiagonal. 

The algebra $\U_a$ is non-simple since the existence of periodic points in $S_a$ implies that $\widehat{\a}$ is 
not minimal.

The algebra $\U_a$ is residually finite: 

When $a\neq1$ is algebraic, we follow the construction of finite dimensional representations 
\cite{ST,Yamashita}: Let $z_q\in \mathrm{Per}_q(S_a)$ be a $q$-periodic point of $\widehat{\a}$. Let 
$\rho_{z_q}: C(S_a) \rightarrow M_{q}(\C)$ be a representation of $C(S_a)$ defined by
$$
\rho_{z_q}(f):=\text{Diag} \big( \, f(z_q),\cdots,f\big(\widehat{\a}^{q-1}(z_q)\big) \,\big) \in M_{q}(\C)
$$
and for $x\in \T$, let
$$
u_{x,z_q}:=\left(\begin{array}{cc} 0&x \\ 1_{q-1}&0 \end{array}\right) \in M_q(\C).
$$
This is a unitary which satisfies the covariance relation 
$u_{x,z_q}^* \, \rho_{z_q}(f) \,u_{x,z_q}=\rho_{z_q}(f \circ \widehat{\a})$
thus $\pi_{x,z_q}:=\rho_{z_q} \rtimes u_{x,z_q}$ is a representation of $\U_a$ on $M_q(\C)$. Again, 
$\pi_x:=\oplus_{q=1}^\infty \oplus_{z_q \in \mathrm{Per}_q (S_a)} \, \pi_{x,z_q}$ is a representation of $\U_a$. 
So, for a dense family $\set{x_l}_{l=1}^\infty$ in $\T$, using the canonical faithful conditional expectation 
$C(S_a) \rtimes_{\widehat{\a}} \Z \rightarrow C(S_a)$ and the density of periodic points, one show that 
$\pi:=\oplus_{l=1}^\infty \pi_{x_l}$ is a faithful representation of $\U_a$ such that 
$\pi(\U_a) \subset \oplus_{l=1}^\infty \oplus _{q=1}^\infty \oplus _{z \in \mathrm{Per}_{q}(S_a)} M_{q}(\C)$.

When $a$ is transcendental, $B_a=\Z \wr \Z$ is an amenable group which has a separating family of finite 
dimensional unitary representations (it is a maximally almost periodic group) since it is residually finite and 
finitely generated, so $\U_a$ is residually finite by \cite[Theorem 4.3]{BL}.

Since the conjugacy classes of $B_a$ are infinite, the von Neumann algebra generated by its left regular
representation is a type $\mathrm{II}_1$ factor (\hspace{-0.3 em} \cite[13.11.14.e]{Dixmier} or 
\cite[III.3.3.7 Prop]{Blackadar}).

$(iii)$ As in \cite[Lemma 2.1]{MV}, the claim on the connected spectrum is equivalent to the conjecture of 
idempotents for $\U_a$. This in turn is a consequence of the Baum--Connes conjecture which has been 
proved by Higson--Kasparov \cite{HK} for torsion-free solvable groups like $G_a=B_a \rtimes_\a \Z$.

$(iv)$ By \cite[Theorem 1.1]{ST}, this property is equivalent to the fact that $\widehat{\a}$ is topologically free, 
namely the aperiodic points in $S_a$ are dense, property already proved in Proposition \ref{prop-dynsyst}.
\end{proof}

A main point of this theorem which is crucial for the sequel is that the algebra $\U_a$ is residually finite.

\subsection{On the classification of algebras $\U_a$}

The algebras $\U_a$ are characterized by their underlying dynamical system:

\begin{theorem}
\label{classification}
Let $\omega_0 \in \R$ and $\kappa \in \R^*_+$ defining $a\neq 1$ in \eqref{defa}. 

(i) $\U_a \simeq \U_{a'}$ yields $c_q(a)=c_q(a'),\, \forall q\in \N^*$.

(ii) The entropy $h(\widehat{\a})$ is also an isomorphism-invariant of $\U_a$.
\end{theorem}

\begin{proof}
$(i)$ The set $\set{c_q(a) \, : \, q \in \N^*}$ defined in \eqref{cq} is an invariant of $C(S_a)\rtimes_{\widehat{\a}} \Z$ 
by \cite{KTW}. This can be also seen concretely in the construction of the above residually finite 
representation $\pi$ since the $\pi_{x,z_q}$ depends only on $c_q$. 

$(ii)$ This is a consequence of $(i)$; see \cite[Theorem 2]{BJ}.
\end{proof}

This result has important physical consequences since a full Lebesgue measure dense set of different parameter 
$a$ (namely the transcendental ones) generates the same algebra or  $\kappa$-deformed space, while in the 
rational case, these spaces are different:

\begin{corollary}
\label{classification1}
As already seen, $\U_a\simeq \U_{1/a}$. Moreover, 

(i) All transcendental numbers $a$ generate isomorphic algebras $ \U_a$.

(ii) If $\U_a\simeq \U_{a'}$, then $a$ and $a'$ are both simultaneously algebraic or transcendental numbers.

(iii) If $\U_a\simeq \U_{a'}$, then $a'=a$ or $a'=a^{-1}$ in the following cases: $a$, $a'$ or their inverses are in 
$\Q^*$ or are quadratic algebraic numbers.

(iv) If $a=m/l \in \Q^*_+$, $K_0(\U_a) \simeq \Z$ and $K_1(\U_a) \simeq \Z \oplus \Z_{l-m}$.

(v) Let a be a quadratic algebraic number with $Q_a=\sum_{j=0}^2 q_jx^j$ and let $l$ be the least common 
multiple of $q_0$ and $q_2$. For $n$ integer, define $n {:} l = n(n, l^{m_0})^{-1}$ where $m_0$ is the maximum 
multiplicity of any prime dividing n and $(n,m)$ is the greatest common divisor of $n$ and $m$ (so $n {:}  l$ is 
obtained by removing from n any prime also dividing $l$).

\qquad If $q_2 \neq q_0$, then $K_0(\U_a) \simeq \Z \oplus \Z_{(q_2-q_0): l}$ and 
$K_1(\U_a) \simeq \Z \oplus \Z_{q_0+q_1+q_2}$.	

\qquad If $q_2 = q_0$, then $K_0(\U_a) \simeq \Z \oplus \Z[\frac{1}{l}]$ and 
$K_1(\U_a) \simeq \Z \oplus \Z[\frac{1}{l}] \oplus \Z_{q_0+q_1+q_2}$.	
\end{corollary}

\begin{proof}
$(i)$ This is proved in section \ref{transcendent}.

$(ii)$ We saw that the entropy $h(\widehat{\a_a})$ is finite if and only if $a$ is algebraic.

$(iii)$ The proof follows from \cite{Brenken1}, but let us give an easy argument in the first case: we may assume 
$a=m/l$ and $a'=m'/l'$ are irreducible rational numbers with $0<m<l$ and $0<m'<l'$.  For $a=m/l$, 
$Q_a(x)=mx-l$,  so $h(\widehat{\a_{m/l}})=\log l$ and $c_1(\widehat{\a_{m/l}})=l-m$. Using Theorem 
\ref{classification} $(i)$, we obtain $l=l'$ and $m=m'$.

$(iv)$ and $(v)$ See \cite{Brenken2}.
\end{proof}
A natural question, turning up in the algebraic case is whether $K$-theory can be used to classify the 
corresponding algebras. However, using the results of \cite{Brenken2}, where algebraic numbers of degree less 
than 3 were considered, we already know that the $K$-groups do not give a complete 
classification: if $a=\sqrt{3}$ and $a'=\sqrt{7}-2$, then $\U_a$ and $\U_{a'}$ are not isomorphic by $(iii)$, but yield  
the same $K$-groups by $(v)$ since $Q_a(x)=x^2-3$ and $Q_{a'}(x)=x^2+4x-3$.

Moreover, Brenken \cite[Proposition 3.1]{Brenken2} also proved the following:

\begin{prop}
If $\tau$ is a tracial state on $\U_a$, then the range of $\tau$ on $K_0(\U_a) $ is $\Z$.
\end{prop}

\subsection{On some representations of $\U_a$ for algebraic $a$}
\label{representationsinduced}

In order to construct spectral triples on $\U_a$ in section~\ref{exist}, it is convenient to get a clear picture of 
some of the essential representations of algebras $\U_a$. We will concentrate on the algebraic case.

In the proof of Theorem~\ref{Thealgebra}, we introduced an irreducible finite dimensional representation 
$\pi_{x,\chi}=\rho_{\chi} \rtimes u_{x,\chi}$ of $\U_a$ on $\C^q$ associated to a $q$-periodic point 
$\chi \in \widehat B_a (= S_a)$ and a $x \in \T$. This representation can be extended to a representation 
$\pi_\chi$ on the Hilbert space $\H_\chi = L^2(\T) \otimes \C^q$ by 
\begin{equation}
\label{eq-reppichi}
\pi_\chi = \int_{\T} \pi_{x,\chi}\, dx.
\end{equation}

In order to get an explicit formula for this representation, denote by $\{e^{(q)}_s \}_{s=1, \dots q}$ the canonical 
basis of $\C^q$ and by $e_n : \theta \mapsto e^{in\theta}$, for $n \in \Z$, the natural basis of 
$L^2(\T) \simeq \ell^2(\Z)$. Then one defines, for any $f \in C^\ast(B_a) = C(\widehat B_a)$:
\begin{align*}
\pi_\chi(f) (e_n \otimes e^{(q)}_s) &= f \circ \widehat{\a}^{s-1}(\chi) \, e_n \otimes e^{(q)}_s, \\
U (e_n \otimes e^{(q)}_s) &= 
\begin{cases}
e_n \otimes e^{(q)}_s & \text{for $1\leq s < q$},\\
e_{n+1} \otimes e^{(q)}_1 & \text{for $s = q$},
\end{cases}
\end{align*}
where $U$ is the generator of $\Z$. This representation is obviously constructed from the representation of 
$G_a=B_a \rtimes_\a \Z$ given by 
$\pi_\chi(b) (e_n \otimes e^{(q)}_s) = \chi \circ \a^{s-1}(b) \, e_n \otimes e^{(q)}_s$ for any $b \in B_a$ (the 
generator $U$ of the action of $\Z$ is the same).

Another natural representation to consider, is the representation of $\U_a$ obtained from the left regular 
representation of $G_a$.

In \cite{LPT} and more systematically in \cite{DJ}, the induced representations \textit{\`a la} Mackey of $G_a$ for 
algebraic $a$ have been investigated. The main results are the following.

For any $\chi \in \widehat B_a$, let the space of functions 
$\varphi : B_a \rtimes_\a \Z \rightarrow \C$ such that $\varphi(b,k) = \chi(b) \varphi(0,k)$ for any $b \in B_a$ and 
$k \in \Z$ be endowed with the norm $\| \varphi \|_\chi^2 = \sum_{k \in \Z} |\varphi(0,k)|^2$. This defines a Hilbert 
space denoted by $\H_\chi^{\mathrm{Ind}}$. The induced representation of $G_a$ on $\H_\chi^{\mathrm{Ind}}$ is 
given by $(\pi_\chi^{\mathrm{Ind}}(g) \varphi)(h) = \varphi(hg)$ for any $g,h \in G_a$. 

It is shown in \cite[Theorem~4.2]{DJ} that this representation is unitarily equivalent to the following one: the 
Hilbert space is $\ell^2(\Z)$ and for any $\xi = (\xi_k)_{k \in \Z}$, one defines 
$(D_\chi(b)\xi)_k := \chi\circ \a^{k}(b) \xi_k$ and the generator of $\Z$ is $(U \xi)_k = \xi_{k+1}$. As a 
representation of $\U_a$, one has $(D_\chi(f)\xi)_k = f \circ \widehat{\a}^{k}(\chi) \xi_k$ for any $f \in C^\ast(B_a)$.

\begin{theorem}
\label{thm-inducedrepresentations}
Assume $a\neq 1$ is algebraic.

(i) There is a natural bijection between the set of orbits of $\widehat{\a}$ in $S_a$ and the set of all equivalence 
classes of induced representations of $\U_a=C^\ast(G_a)$. This bijection is realized by 
$\chi \mapsto \pi_\chi^{\mathrm{Ind}}$.

(ii) The representation $\pi_\chi^{\mathrm{Ind}}$ is irreducible if and only if $\chi$ is aperiodic.

(iii) The commutant of $\pi_\chi^{\mathrm{Ind}}$ for a $q$-periodic point $\chi$ is the commutative algebra $C(\T)$.

(iv) The right regular representation $R$ of $\U_a=C^\ast(G_a)$ is unitarily equivalent to the representation
\begin{equation*}
\int_{B_a}^{\oplus} \pi_\chi^{\mathrm{Ind}} \, d\mu(\chi).
\end{equation*}
\end{theorem}

\begin{proof}
These results are proved in \cite{DJ} for the corresponding representations of the group $G_a$: $(i)$  is 
Theorem~4.13, $(ii)$ and $(iii)$ are Theorem~4.8 and $(iv)$ is Theorem~6.3.
\end{proof}

For a $q$-periodic $\chi$, the representation $\pi_\chi^{\mathrm{Ind}}$ is reducible. Explicitly one has:

\begin{prop}
If $\chi$ is $q$-periodic, then $\pi_\chi^{\mathrm{Ind}}$ is unitarily equivalent to the representation 
$\pi_\chi$ on $\H_\chi$, so that its continuous decomposition into irreducible finite dimensional representations 
on $\C^q$ is realized by \eqref{eq-reppichi} along $\T$.
\end{prop}

\begin{proof}
$\pi_\chi^{\mathrm{Ind}}$ is unitarily equivalent to the representation $D_\chi$ on $\ell^2(\Z)$. The unitary map 
between the two Hilbert spaces $\H_\chi$ and $\ell^2(\Z)$ defined by the correspondence 
$e_n \otimes e^{(q)}_s \mapsto \delta_{nq+r-1}$ with $(\delta_k)_{k \in \Z}$ as natural basis of $\ell^2(\Z)$, 
intertwines the representations $\pi_\chi$ and $D_\chi$.
\end{proof}

This proposition states that, while the finite dimensional representations of $\U_a$ are not obtained as induced 
representations, they are nevertheless reductions of induced representations. The right regular representation 
$R$ contains the infinite dimensional irreducible induced representations which are only accessible using 
aperiodic points. The representations $R$ and $\pi_\chi$ are not quasi-equivalent : this difference will play a 
crucial role in the construction of different spectral triples, see Remark \ref{Contradict}.

Despite the fact that the $\pi_\chi^{\mathrm{Ind}}$'s yield a von Neumann factor of type $\mathrm{I}$, $R$ gives 
a type $\mathrm{II}_1$ factor because of the integral. So the group $G_a$ is non-type $\mathrm{I}$.

\section{The particular case $a=m \in \N^*$}
\label{abstractapproach}

According to Lemma \ref{symmetry}, the case $a=m \in \N^*$ considered in this section also covers the case 
$a=1/m$. We do insist on this special compact version of a $\kappa$-deformed space since in this case the algebra 
is generated by two unitaries related by one relation (see \eqref{def} below) in the spirit of the popular 
noncommutative two-torus: concretely, $G_m=BS(1,m)$ is the Baumslag--Solitar group which is generated by two 
elements with a one-relator while, when $a$ and $a^{-1}$ are not integers, $G_a$ is not a finitely presented group 
(even if it has two generators). Thus this assumption simplifies the computations of section \ref{algebrique}.

Moreover, the results described in this section rely more on some well-known properties of the Baumslag--Solitar 
group than on the dynamical system used until now. As a consequence, these results (which for the most are already  
valid and exposed for generic values of $a$) are presented and proved independently. These structures appears 
also naturally in wavelet theory, which could benefit from our analysis.

\subsection{The algebra}
\label{thealgebra}

\begin{definition}
\label{def-thealgebra}
Let $\U_m$ be the universal $C^*$-algebra labelled by $m \in \Z^*$ (restricted to $\N^*$ later) and 
generated by two unitaries 
$U$ and $V$ such that  
\begin{align}
\label{def}
UVU^{-1}=V^m.
\end{align}
\end{definition}
This universal $C^*$-algebra $\U_m$ is denoted by $\OO(E_{1,m})$ in \cite{Kat}, and also $\OO_{m,1}(\T)$ in 
\cite{Yamashita} (where only $m\in\N^*$ is considered.) These algebras are topological graph $C^*$-algebras 
which can be seen as transformation group $C^*$-algebras on solenoid groups as already noticed in 
\cite{BJ,J,Brenken}. They have been used in wavelets and coding theory \cite{DHPQ,DJ,DJP}. 

Relation \eqref{def} also appeared in the Baumslag--Solitar group $BS(1,m)$ introduced in \cite{BS} as the group 
generated by two elements $u,\,v$ with a one-relator:
$$
BS(1,m):=\langle u,\,v \,\vert \, uvu^{-1}=v^m \rangle.
$$
This group plays a role in combinatorial and geometric group theory. It is a finitely generated, meta-abelian, 
residually finite, Hopfian, torsion-free, amenable (solvable non-nilpotent) group. It has infinite conjugacy classes,  
a uniformly exponential growth (for $m\neq 1$) but is not Gromov hyperbolic \cite{Harpe}. 
There exists a monomorphism $\pi$ from $BS(1,m)$ into the  $PSL(2,\R)$ group (see 
also~\eqref{eq-repGaGL2R}) given by
$$
\pi(u):= \left(\begin{array}{cc} \sqrt{m}&0 \\ 0&\sqrt{m^{-1}} \end{array}\right), \quad 
\pi(v):= \left(\begin{array}{cc} 1&1 \\ 0&1 \end{array}\right).
$$
Note that $BS(1,1)$ is the free abelian group on two generators and $BS(1,-1)$ is the Klein bottle 
group.

As for the $BS(1,m)$ groups, within the algebras $\U_m$, we remark that $\U_1$ and $\U_{-1}$ play a 
particular role: $\U{_1} = C(\T^2)$ (continuous functions on the 2-torus), and  
$\U_{-1}\supset C(\T^2)$ will not be considered here since we need $a=m>0$.

For $m\geq 2$, a solenoid appears as in section~\ref{algebrique}, as well as a crossed product structure, a fact that 
we recall now in this particular context.

Assume $m\geq 2$ and let the subring of $\Q$ generated over $\Z$ by $\tfrac{1}{m}$
$$
B_m=B_{1/m}:=\Z [\tfrac{1}{m}]:=\bigcup_{l\in \N}m^{-l} \Z \subset \Q.
$$
It is the additive subgroup of $\Q$ which is an inductive limit of the rank-one groups $m^{-l}\Z$, for 
$l=0,1,2,\dots$ and $B_m$ has a natural automorphism $\a$ defined by 
$$
\a(b):=mb.
$$
Note that the abelian group $B_m$ is not finitely generated. When $m\rightarrow \infty$, 
$BS(1,m) \rightarrow \Z \wr \Z $ (in the space of marked groups on two generators) \cite{Stalder}. This group also 
appears when $m=e^{-\omega_0/\kappa}$ is replaced by a transcendental number $a\in \R^*_+$ as seen in section 
\ref{aquelconque}.

$B_m$ can be identified with the subgroup of the affine group Aff$_1(\Q)$ generated by the dilatation 
$u:x \rightarrow mx$ and the translation $v: x\rightarrow x+1$. It is the subgroup normally generated in $BS(1,m)$ 
by $v$,  $\langle v: u^{-1} vu, u^{-2} v u^2 , . . .\rangle$.
The Baumslag--Solitar group $BS(1,m)$ is then isomorphic to the crossed product 
$$
BS(1,m) \simeq B_m \rtimes_\a \Z
$$
so that one has the group extension 
$$
1\rightarrow  \Z[ \tfrac{1}{m}] \rightarrow B(1, m) \rightarrow \Z \rightarrow 1. 
$$

Using this crossed product decomposition, the group $BS(1,m)$ has the following explicit law: 
$(b,l)(b',l')=(b+\a^l(b'),l+l')$ for $l,l'\in \Z$ and $b,b'\in B_m$. It is of course generated by the elements $u:=(0,1)$ 
and $f_b:=(b,0)$ with $b\in B_m$. These generators satisfy for $j,l\in \Z, n\in\N$
\begin{align}
\label{gen}
uf_bu^{-1}=f_{\a(b)}, \text{ and if } 
f_{\a^{-n}}j:=(\a^{-n}j,0)=u^{-n}f_j u^n, \text{ then } ((\tfrac{1}{m})^{n}j,l)=f_{{(\tfrac{1}{m})}^nj}\,u^l.
\end{align} 

$BS(1,m)$ is a subgroup of the $``ax+b"$ group (with law 
$(b,a)(b',a'):=(b+ab',aa')$) and can be viewed as the following subgroup of two-by-two matrices 
$\set{ \left(\begin{array}{cc} m^l&b \\ 0&1 \end{array}\right) \; : \; l\in \Z,\,b\in B_m}$.  Note that 
$BS(1,m)\simeq B(1,m')$ is equivalent to $m=m'$ \cite{Mol}.

It is interesting to quote that  if $p_1,\cdots,p_n$ are the prime divisors of $m$, then the diagonal embedding 
$BS(1,m) \hookrightarrow {\rm Aff}_1(\Q_{p_1}) \times \cdots \times {\rm Aff}_1(\Q_{p_n}) \times {\rm Aff}_1(\R)$ 
has a discrete image. 

Let $\widehat {B_m}$ be the Pontryagin dual of $B_m$ endowed with the discrete topology. It is isomorphic to the 
solenoid
\begin{align}
\label{solenoid}
S_m =S_{1/m}\simeq \set{(z_k)_{k=0}^{\infty} \in \prod_{i=0}^\infty \T \, : \, z_{k+1}^m=z_k, \,\forall k\in \N_0}
\end{align}
using $z_k:=\chi\big((\tfrac{1}{m})^k\big)$ for any $\chi \in \widehat B_m$. The group $S_m$ is compact 
connected and abelian. Notice that (see section~\ref{algebrique}) 
$S_m \simeq \set{(z_k)_{k=-\infty}^{\infty} \in  \T^\Z \, : \, z_{k+1}^m=z_k, \, k\in \Z}$ 
defining $z_{-k}:=z_0^{mk}$ for $k>0$.

The natural embedding 
$$
\hat{\iota} : \theta \in\R \mapsto \chi_\theta \in S_m \text{ where }\chi_\theta(b):=e^{i2\pi \theta b} \in \T 
\text{ for } b\in B_m
$$
identifies $S_m$ as the Bohr compactification $b_{B_m}\R$ of $\R$.

$S_m$ is endowed with a natural group automorphism $\widehat\a$ given by 
$$
\widehat{\a} (z_0,z_1,z_2,\dots)=(z_0^m,z_0,z_1,\dots) 
\qquad\text{ and }\qquad \widehat{\a} ^{-1}(z_0,z_1,z_2,\dots)=(z_1,z_2,\dots).
$$

There are periodic points in $S_m$ of arbitrary period $q\in \N^*$ and they are all of the 
following form: if $z_0$ is a solution of $z^{m^{q}-1} =1$, then 
$(z_0,z_0^{m^{q-1}},\dots,z_0^m,z_0,\dots) \in S_m$; so there are only finitely many periodic points, namely 
$c_q(m) = m^{q}-1$ such points.

The $C^*$-algebra $C(S_m) \simeq C^*(B_m)$ is precisely the algebra of almost periodic 
functions on $\R$, with frequencies in $B_m$ and the isomorphism is the map 
$f \mapsto f \circ \hat\iota$. 

It follows that for the $C^*$-algebra we consider, one has
$$
\U_m = C^*(BS(1,m)) \simeq C^*(B_m)\rtimes_{\a} \Z \simeq C(S_m) \rtimes_{\widehat\a} \Z.
$$
The unitary element $U$ of Definition~\ref{def-thealgebra} is precisely the generator of the action $\a$ of $\Z$ on 
$C^*(B_m)$ while $V$ is one of the generators $\set{ U^{-\ell}V U^\ell : \ell \in \Z}$ of the abelian 
algebra $C^*(B_m)$. As a continuous function on $S_m$, $U^{-\ell}V U^\ell$ is the function 
$(z_k)_{k=0}^\infty \mapsto z_\ell$ and in particular, $V:(z_k)_{k=0}^\infty \mapsto z_0$.

\vspace{0.3cm}

Note that the subgroup $\set{z:=(z_k)_{k=0}^\infty \in S_m \, : \, \widehat\a^q (z)=z \text{ for some } 
q\in \N^*}$ of periodic points is dense in $S_m$ and $\widehat\a$ is ergodic on $S_m$ for $m\geq 2$ as previously 
seen \cite[Proposition 1]{BJ}.

\subsection{The representations}

The knowledge of $^\ast$-representations of $\U_m$ is important, and even essential in the context of spectral 
triples (see Definition \ref{defi}). According to \eqref{gen}, any unitary representation of $BS(1,m)$ is given by a 
unitary operator $U$ and a family of unitaries $T_k$, $k\in \Z$, with the constraint $UT_kU^{-1}=T_{mk}$, so there 
is a bijection between the $^\ast$-representations of $\U_m$ on some Hilbert space $\H$ and the corresponding 
unitary representations of $BS(1,m)$. This is rephrased usefully in the following lemma \cite{J}:

\begin{lemma}
\label{useful}
The algebra $\U_m$ is the $C^*$-algebra generated by $L^\infty(\T)$ and a unitary symbol 
$\tilde U$ with commutation relations
\begin{align}
\tilde U \,f\, \tilde U^{-1}=f \circ e_m, \,\,\forall f \in L^{\infty} (\T)
\end{align}
where $e_n(z):=z^n$.
\end{lemma}

\begin{proof}
A $^\ast$-representation $\pi$ of $\U_m$ on some arbitrary Hilbert space $\H$ can be obtained from a 
couple ($\tilde\pi,\tilde U)$ with  a $^\ast$-representation $\tilde \pi$ of $L^\infty(\T)$ and a unitary 
operator $\tilde U$ on $\H$ related by  $\tilde U \tilde \pi(f(z)) \tilde U^{-1}=\tilde \pi \big(f(z^m)\big)$ 
and $\tilde V:=\tilde\pi(e_1)$ via $\pi(U):=\tilde U$ and $\pi(V):=\tilde V$. Conversely, given a 
representation $\pi$ of $\U_m$, define $\tilde \pi(f):=f\big(\pi(V)\big)$ via the spectral theorem, 
for any $f\in L^{\infty}(\T)$, and $\tilde U:=\pi(U)$, so \eqref{def} implies 
$\tilde U \tilde \pi(f(z)) \tilde U^{-1}=\tilde \pi \big(f(z^m)\big)$.
\end{proof}

In particular, $\U_m$ contains a family of abelian subalgebras 
$\A_n:=\tilde U^{-n}\, L^\infty(\T)\,\tilde U^n$ for $n\in \N$, which is increasing since 
$\tilde U^{-n} \,f \,\tilde U^n=\tilde U^{-(n+1)} \,f\circ e_m \, \tilde U^{(n+1)}$.

Remark that $\set{\sum_{n\in \N, \, {\rm finite}} \,f_n \tilde U^n+g_n \tilde U^{-n} \, : \, f_n,g_n \in L^\infty(\T)}$ 
is a dense *-subalgebra of $\U_m$.
\vspace{0.5cm}

If we choose the Hilbert space $\H:=L^2(\R)$, the scaling and shift operators give rise to a 
representation $\pi$ of $ \U_m$ on $\H$ by $\pi(U): \psi(x) \mapsto \tfrac{1}{\sqrt{m}} \, 
\psi(\tfrac{x}{m})$ and $\pi(V): \psi(x) \mapsto  \psi(x-1)$. 

The Haar measure $\nu$ on $\widehat B_m$ gives rise to a faithful trace on $\U_m$ and, since there are many 
finite dimensional representations of $\U_m$ (see proof of Theorem \ref{Thealgebra}), there are many traces on it.

If we choose $\H:=L^2(S_m,\nu)\simeq \ell^2(B_m)$, then $C(S_m)$ acts on $\H$ by left 
pointwise multiplication and we define $U: \psi \in \H \mapsto \psi \circ \widehat\a \in \H$ and get a 
covariant representation of $(S_m,U)$ of the dynamical system $(C(S_m),\widehat{\a},\Z)$, so a 
representation of $\U_m$ on $\H$. Since $\widehat\a$ is ergodic, this representation is irreducible and 
faithful \cite[Theorem 1]{BJ}.

If we choose $\H:=\ell^2(\Z)$, then for each $\theta \in \R$, we get an induced representation of 
$\U_m$ by 
$$
\pi_\theta(U)\psi(k):=\psi(k-1) \text{ and }\pi_\theta(V)\psi(k):=\chi_\theta( m^{-k})\,\psi(k), 
\text{ for }k\in \Z.
$$

\bigskip
We summarize a few results on the algebras $\U_m$ (independently of Theorem 
\ref{Thealgebra}).

\begin{prop}  
\label{casentier}
Let $m\in \N$, $m\geq 2$.

(i) $\U_m$ is a NGCR, $AF$-embeddable, non-simple, residually 
finite dimensional $C^*$-algebra and its generated von Neumann algebra for the left representation is a type 
$\mathrm{II}_1$-factor.

(ii) $\big(K_0(\U_m),[1_{\U_m}],K_1(\U_m)\big) \simeq (\Z,1,\Z \oplus \Z_{m-1})$.

(iii) $\U_m \simeq \U_{m'}$ if and only if $m=m'$.
\end{prop}

\begin{proof}
$(i)$: see \cite{Brenken,BJ} and \cite[Theorem 2.2]{Yamashita},\cite{Ivanov}.

$(ii)$: see \cite{Brenken2} or \cite[Example A6]{Kat}.

$(iii)$: consequence of $(ii)$ (or Corollary \ref{classification1}).
\end{proof}

\section{On the existence of spectral triples}
\label{exist}

Since we want to construct spectral triples on $\U_a$, it is worthwhile to know the heat decay of 
$G_a=B_a\rtimes_\a \Z$ via a random walk on the Cayley graph of $G_a$ with generators 
$S=\set{x,x^{-1},y,y^{-1}}$ where $x=(0,1)$ and $y=(1,0)$, and with a constant weight and standard Laplacian. 

The decay of the heat kernel $p_t$, with $t\in \N$, has been computed on the diagonal in 
\cite[Theorem 1.1]{Pittet}, \cite[Theorem 5.2]{CGP}: when $t \rightarrow \infty$, 
$p_{2t} \sim e^{-t^{1/3}\,(\log t)^{2/3}}$ if $a$ is transcendental while $p_{2t} \sim e^{-t^{1/3}}$ if $a$ is 
algebraic. This is related to the fact that $G_a$ has exponential volume growth.

However, for a finite dimensional connected non-compact Lie group, the behaviour of the heat kernel $p_t$ 
depends on $t\in\R^*_+$ and can diverge for the short time behaviour when $t\rightarrow 0$. 
Let us explain how this point is related to the dimension:

In noncommutative geometry, a (regular simple) spectral triple $(\A,\H,\DD)$ has a (spectral) dimension 
which is given by 
$\max \set{n\in \N \, : \, n \text{ is a pole of } \zeta_\DD:s \in \C \rightarrow \Tr(\vert \DD \vert ^{-s})}$ (here 
$\DD$ is assumed invertible). 
In particular, when $M$ is a $n$-dimensional compact Riemannian spin manifold, and $\A=C^\infty(M)$, 
$\H=L^2(S)$ where $S$ is the spinor bundle and $\DD$ is the canonical Dirac operator, the spectral 
dimension coincides with $n$. Via the Wodzicki residue, an integral 
$\ncint X:=\Res_{s=0} \Tr(X \vert \DD \vert^{-s})$ is defined on (classical) pseudodifferential operators $X$ 
acting on the smooth sections of $S$. 
For instance, $\ncint \vert \DD \vert^{-n}$ coincides (up to a universal constant) with the Dixmier trace 
$\Tr_{\text{Dix}}(\vert \DD \vert ^{-n})=\lim_{N\rightarrow \infty} \log(N)^{-1}\sum_{k=1}^N \vert \lambda_k 
\vert ^{-n}$ where the $\lambda_k$ are the singular values of $\DD$. The dimension of $M$ appears in 
$\Tr(e^{-t\DD^2}) \sim \sum_{N \geq 0}\tfrac{1}{t^{(n-N)/2}} \,a_N(\DD)$ when $t \rightarrow 0$. In particular, 
when $M=\R^n$ with Lebesgue measure and $\DD^2=-\triangle$ is the standard Laplacian 
(non-compactness is not a problem), the heat kernel is $p_t(x,x)=\tfrac{1}{(4 \pi t)^{n/2}}$ for all $x \in M$ 
(see \cite{Con94,ConnesMarcolli,Polaris}). As a consequence, $\Tr(\vert \DD \vert^{-(n+\epsilon)}) < \infty$ 
for all $\epsilon >0$. 

We will see in this section that, depending on the chosen representation of $\U_a$, such an $n$ does not always 
exist, meaning that the ``dimension is infinite".

\begin{definition}
\label{defi}
A spectral triple (or unbounded Fredholm module) $(\A,\H,\DD)$ is given by a 
unital $C^*$-algebra $A$ with a faithful representation $\pi$ on a Hilbert space $\H$ and an unbounded
self-adjoint operator $\DD$ such that 

\qquad - the set $\A=\set{a \in \A \,: \, [\DD, \pi(a)] \text{ is bounded }}$ is norm dense in $A$,

\qquad - $(1+\DD^2)^{-1} \in J$ where $J$ is a symmetrically-normed ideal of the compact operators 
$\K(\H)$ on $\H$.

(Note that $\A$ is always a $^*$-subalgebra of $A$.)

The triple is $p$-summable if $J=\L^p(\H)$ for $1 \leq p<\infty$ (i.e. $\Tr\big((1+\DD^2)^{-p/2} \big) <\infty$.

It is $p^+$-summable if $J=\L^{p+}(\H)$.

It is finitely summable if it is $p$-summable for some $p$.

It is $\theta$-summable if there exists $t_0\geq 0$ such that $\Tr\big(e^{-t\DD^2} \big) <\infty$ for all $t>t_0$  
(thus $J=\K(H)$).
\end{definition}
Note that $p$-summability implies $\theta$-summability.

Connes proved in \cite{Con89} that, for an infinite, discrete, non-amenable group $G$, there exist no 
finitely summable spectral triples on $A=C^*_{red}(G)$. However,  in this case, there always exist 
$\theta$-summable spectral triples on $A$ (even with $\DD>0$). 
Using a computable obstruction to the existence of quasicentral approximate units relative to $J$ for $A$, 
Voiculescu was able to derive, for solvable groups with exponential growth, the non-existence result for 
unbounded (generalized) Fredholm modules using the Macaev ideal $J=\L^{\infty,1}(\H)$ \cite{Voiculescu}. 
We use these results in the following:

\begin{theorem} 
\label{nonexistence}
Non-existence of finite-summable spectral triples.

Let $ A=\U_a$, $G_a=B_a \rtimes_\a \Z$ and $\A=\C [G_a]$.

(i) There is no finitely summable spectral triple $\big(\pi(\A), \H_\pi,\DD\big)$ when the representation $\pi$ 
is quasiequivalent to the left regular one.

(ii) There exist $\theta$-summable spectral triples $\big(\pi(\A), \H_\pi,\DD\big)$ with $t_0=0$ where the 
representation $\pi$ is quasiequivalent to the left regular one.
\end{theorem}

\begin{proof}
$(i)$ Since $\U_a \simeq C^*_{red}(G_a)$ and $G_a$ is a solvable group with 
exponential growth, this follows from \cite[4.12. Prop]{Voiculescu}.

$(ii)$ Such a triple is given in \cite[Prop. 24]{Con89} where the representation $\pi$ is the left regular one on 
$\H_\pi=\ell^2(G_a)$ and $\DD:=\DD_\ell$ is the multiplication by the length function $\ell$ 
on the group $G_a$ associated to the generating set $\set{u,v,u^{-1},v^{-1}}$ where $u=(0,1)$ and $v=(1,0)$ (see section~\ref{aquelconque}). In particular 
$\DD_\ell>0$.
\end{proof}

Despite the previous result, we add a few explicit examples of spectral triples using the fact that the 
algebra $\U_a$ is residually finite. Clearly, these triples deal with a restrictive part of the geometry of the 
$\kappa$-deformation based on $\U_a$, namely the dynamical system which is behind. The residually 
finite property is seen via the periodic points of this dynamics.

We gather in the following theorem all these results:

\begin{theorem} 
\label{existence}
Existence of finite-summable spectral triples.

Let $ A=\U_a$ and $\A=\C [G_a]$.

(i) There exist spectral triples $\big(\pi(\A), \H_\pi,\DD\big)$ which are compact, i.e. $[\DD,\pi(x)]$ is compact 
for all $x \in \A$.

(ii) There exist spectral triples $\big(\pi(\A), \H_\pi,\DD\big)$ such that $[\DD,\pi(x)]=0, \, \forall x \in \U_a$, 
and with arbitrary summability.

(iii) When $a$ is algebraic, there exist spectral triples $\big(\pi(\A), \H_\pi,\DD\big)$ such that 
$[\DD,\pi(v)]=0$, $[\DD,\pi(u)] \neq 0$ and with an arbitrary summability $p\geq 2$.
\end{theorem}

In case $(i)$, $[\DD,\pi(x)]$ is not necessarily zero but the summability is not controlled while for case $(ii)$, 
the condition $[\DD,\pi(x)]=0$ enables us to control summability. In a sense, case $(iii)$ is a mixed situation 
requiring that $a$ be algebraic. In that situation, we have a concrete description of the representation $\pi$ so that 
an explicit formula for the Dirac operator can be proposed.

\begin{proof}
$(i)$ Since $\U_a$ is a residually finite $C^*$-algebra (i.e. faithfully embeddable into the algebra 
$\prod_{n\in N} M_{k_n}(\C)$), it is quasidiagonal \cite[Prop. 7.1.15]{BO}. Thus the result follows from
\cite[Theorem 2.1]{SZ}: let $\pi$ be a faithful quasidiagonal representation which means that there exists 
an increasing family of finite rank projections $(P_n)_{n\in \N}$ such that $P_n$ strongly converges to the 
identity and $[P_n, \pi(x)] \rightarrow 0$ for all $x \in \U_a$. 
Define $Q_n:=P_n-P_{n-1}$ ($P_0=0$) and choose a sequence $d_n\in \R$, 
$\vert d_n \vert \nearrow \infty$. Select a subsequence $(P_{n_k})_k$ such that 
$\sum_k \vert d_{n_k} \vert \, \vert \vert [Q_{n_k} , x] \vert \vert < \infty$ for $x=u,v$. Then for 
$\DD:=\sum_k d_{n_k} Q_{n_k}$, $[\DD, \pi(x)]$ is compact for $x\in \A$ and $(\pi(\U_a), \H_\pi,\DD)$ is a 
spectral triple. Note that the family $(Q_{n_k})_k$ depends on the sequence $(d_n)_n$, so this latter 
cannot be adjusted to control summability.

$(ii)$ \cite{SZ}: since $\U_a$ is  $^*$-isomorphic to a subalgebra of $\prod_{n\in \N} M_{k_n}(\C)$, there 
exists a faithful representation $\pi$ of $\U_a$ acting on $\H_\pi=\bigoplus_{n \in \N} \C^{k_n}$. Let 
$P_n$ be the projection on $\C^{k_n}$ and as above,  $\DD:= \sum_{n=1}^\infty d_n \,Q_n$. Then 
$[\DD, \pi(x)]=0$ for $x \in A$ (we can even choose $\A=A$) and the summability is arbitrary by choosing 
$(d_n)_n$.

$(iii)$ We first construct a faithful representation of $\U_a$ inspired by \cite{Yamashita}. For any  non zero positive 
integer $q$, we already defined
$$
\mathrm{Per}_q(S_a) = \set{z=(z_k)_{k=0}^\infty \in S_a \, : \,  \widehat{\a}^{k}(z) \neq  z, \ \forall  
k<q, \ \widehat{\a}^{q}( z) =  z}
$$ 
the space of $q$-periodic points in $S_a$ for the action $\widehat{\a}$  
of $\Z$. 

Since $c_q:=\#\mathrm{Per}_q(S_a) = \vert q_d \vert^q \, \Pi_{k=1}^d \vert 1-{r_k}^q \vert< \infty$ when $a$ is 
algebraic, we can enumerate these periodic points as $\{ z^{(i)}\}_{i=1, \dots, c_q} = \mathrm{Per}_q(S_a)$. 

Consider the Hilbert space 
\begin{equation*}
\H := \bigoplus_{q \geq 1} L^2(\T) \otimes \C^q \otimes \C^{c_q}
\end{equation*}
We denote by $E^{(q)}_i \in M_q(\C)$, for $i = 1, \dots, q$ the matrix whose only non-vanishing entry is 
$1$ on the diagonal at the $i$th entry from the top. We denote by $\iota : \T \rightarrow \T$ the identity 
map, which we identify also with $\iota : \R \rightarrow \T$, $\iota(\theta) = e^{i\theta}$. 
Let us introduce a convenient basis in $\H$: denote by $\{e^{(q)}_s \}_{s=1, \dots, q}$, the 
canonical basis of $\C^q$, $\{e^{(c_q)}_i \}_{i=1, \dots, c_q}$, the 
canonical basis of $\C^{c_q}$, and by $e_n : \theta \mapsto e^{in\theta}$, for $n \in \Z$, the natural basis of 
$L^2(\T) \simeq \ell^2(\Z)$. Then, for $n \in \Z$, $q \geq 1$, $r = 1, \dots, q$ and $i = 1, \dots, c_q$, the 
vectors $e_n \otimes e^{(q)}_r \otimes e^{(c_q)}_i$ define an orthonormal basis of $\H$: any $\psi \in \H$ 
can then be decomposed as
\begin{equation*}
\psi = \bigoplus_{n \in \Z} \bigoplus_{q \geq 1} \sum_{r=1}^{q} \sum_{i=1}^{c_q} \,\psi_{n,q,r,i}\ e_n \otimes 
e^{(q)}_r \otimes e^{(c_q)}_i = \bigoplus_{n \in \Z} \bigoplus_{q \geq 1} \sum_{i=1}^{c_q} \,e_n \otimes 
\begin{pmatrix}
\psi_{n,q,1,i} \\ \vdots \\ \psi_{n,q,q,i}
\end{pmatrix}
\otimes e^{(c_q)}_i
\end{equation*}
for $\psi_{n,q,r,i} \in \C$.

We define on $\H$ a faithful representation $\pi$ of $\U_a$ on any $f \in C(S_a)$ and the unitary 
generator $u$ of $\Z$ in $\U_a = C(S_a) \rtimes_{\widehat{\a}} \Z$ by
\begin{align*}
&\pi(f) := \bigoplus_{q \geq 1} \ 1_{L^2(\T)} \otimes \sum_{i=1}^{c_q} \left[
\begin{pmatrix} 
f(z^{(i)}) &  & 0 \\
 & \ddots &  \\
0 & & f \circ \widehat{\a}^{q-1}(z^{(i)})
\end{pmatrix}
\otimes E^{(c_q)}_i \right],\\
&\pi(u) := \bigoplus_{q \geq 1} M_\iota \otimes \begin{pmatrix} 0 & 1 \\ 0_{q-1}&0 \end{pmatrix} \otimes 1_{c_q} + 
1_{L^2(\T)} \otimes \begin{pmatrix} 0 & 0\\ 1_{q-1}&0 \end{pmatrix} \otimes 1_{c_q}
\end{align*}
where $M_\iota$ is the multiplication by the function $\iota$. In particular,
$$
\pi(v):=\bigoplus_{q \geq 1} \ 1_{L^2(\T)} \otimes \sum_{i=1}^{c_q} \left[
\begin{pmatrix} 
{z^{(i)}}_0 &  & 0 \\
 & \ddots &  \\
0 & &  (\widehat{\a}^{q-1}{z^{(i)})}_0
\end{pmatrix}
\otimes E^{(c_q)}_i \right].
$$
We define now a self-adjoint unbounded operators on $\H$. It is taken to be diagonal, with 
eigenvalues given by  a sequence 
$\{ \mu_{n,q, i} \}_{n\in\Z, \,q \geq 1,\, i=1, \dots c_q}$ of real parameters and it is given by:
\begin{equation*}
\DD \,\,\big(e_n \otimes e^{(q)}_r \otimes e^{(c_q)}_i \big)\, :=\, \mu_{n,q, i} \, \, \big(e_n 
\otimes e^{(q)}_r \otimes e^{(c_q)}_i \big).
\end{equation*}

Then, one has the following relations:
\begin{align*}
[\DD, \pi(f)] &= 0, (\text{so } [\DD,\pi(v)]=0),
\\
[\DD, \pi(u)]  \,e_n \otimes e^{(q)}_r \otimes e^{(c_q)}_i &= 
\delta_{r,q} \, (\mu_{n+1,q, i} - \mu_{n,q, i}) \, e_n \otimes e^{(q)}_1 \otimes e^{(c_q)}_i.
\end{align*}

By construction, $\DD$ has well-defined eigenvectors and eigenvalues: 
\begin{equation*}
\H_{n,i}^{(q)} := \text{Span} \set{ e_{n,r,i}^{(q)} := e_n \otimes e^{(q)}_r \otimes e^{(c_q)}_i \, : \, \ r=1, \dots, q }
\end{equation*}
is a $q$-dimensional eigenspace of $\DD$ for the eigenvalue $\mu_{n,q,i}$.

Thus, for any $0<p<\infty$, 
\begin{equation*}
\Tr\big((1+\DD^2)^{-p/2}\big) = 
\sum_{n\in\Z,\, q\geq 1,\,i=1\dots c_q} \frac{q}{{\big(1+(\mu_{n,q,i})^2}\big)^{p/2}}\,.
\end{equation*}

Let us take $\mu_{n,q,i} = \lambda_q(| n | + b_q)$ where $b_q$ does not depend on $n$ and $i$ and 
$\lambda_q = \pm 1$. 
Then $\DD$ is invertible. 

Since $| \mu_{n+1,q, i} - \mu_{n,q, i}| = |\, | n+1 | - | n | \,| = 1$, one proves that $[\DD, \pi(u)]$ is a bounded operator on 
$\H$.

For $p >2$, take $b_q = q (q c_q)^{2/p}$. Then, one has 
\begin{align*}
\sum_{n\in\Z,\, q\geq 1,\,i=1\dots c_q} \frac{q}{{\big(1+(\mu_{n,q,i})^2}\big)^{p/2}}
& =
\sum_{n\in\Z,\, q\geq 1} \frac{q c_q}{{\big(1+(| n | + b_q)^2}\big)^{p/2}} \\
& \leq
\sum_{n\in\Z,\, q\geq 1} \frac{q c_q}{(2 | n | b_q)^{p/2}} \\
& =
\frac{1}{2^{p/2}} \sum_{n\in\Z,\, q\geq 1} \frac{1}{| n |^{p/2}} \frac{1}{q^{p/2}} \\
&< \infty \text{ as a product of two finite sums.}
\end{align*}

For $p = 2$, the previous inequalities are inefficient but one has
\begin{align*}
\sum_{n\in\Z,\, q\geq 1} \frac{q c_q}{1+(| n | + b_q)^2}
&\leq
\sum_{n\in\Z,\, q\geq 1} \frac{q c_q}{(| n | + b_q)^2}\,.
\end{align*}
Moreover, $\sum_{n\geq 0} \frac{q c_q}{(| n | + b_q)^2} = q c_q \zeta(2,b_p)$, where $\zeta(s,z)$ is the generalized 
Riemann zeta function. For $x \to \infty$, $\zeta(2,x) \sim \frac{1}{x}$. Thus, with the choice 
$b_q = q^3 c_q$, the sum $\sum_{q\geq 1} q c_q \zeta(2,b_p)$ is finite, so that $\Tr\big((1+\DD^2)^{-1}\big) < \infty$. 
\end{proof}

\begin{remark}
\label{Contradict}
There is no contradiction between Theorems~\ref{nonexistence} and \ref{existence} since the faithful quasidiagonal 
representation (or residually finite one) $\pi$ of $\U_a$ used above to construct $\DD$ is not quasiequivalent to the 
left regular one: actually, as already mentioned, the von Neumann algebra generated by $\pi(\U_a)$ is a 
$\mathrm{II}_1$ factor when $\pi$ is the left regular representation, while it is of type $\mathrm{I}$ when $\pi$ is the 
quasidiagonal or residually finite one \cite[5.4.3.]{Dixmier}. 

A more direct way to confirm that the representation $\pi$ used in the proof of point $(iii)$ of 
Theorems~\ref{existence} is not quasiequivalent to the left regular representation (or to the right regular 
representation which is unitarily equivalent to the left one) is to notice that $\pi$ is the direct integral
\begin{equation*}
\pi = \int_{\mathrm{Per}}^{\oplus} \pi_\chi \, d\mu(\chi)
\end{equation*}
of the \emph{finite dimensional} representations $\pi_\chi$ defined in \eqref{eq-reppichi}. As such, this 
representation is strictly contained in the right regular representation $R$ as can be checked using $(iv)$ of 
Theorem~\ref{thm-inducedrepresentations}. The part of $R$ which is not in $\pi$ is given by the induced 
\emph{infinite dimensional} irreducible representations constructed on aperiodic $\chi$'s.

As noticed in \cite{SZ}, if $(\A,\H_\pi,\DD)$ is a spectral triple such that $[\DD,\pi(x)]=0, \, \forall x \in A$, 
then A is a residually finite $C^*$-algebra.
\end{remark}

\begin{remark}
Theorem \ref{nonexistence} essentially says that the 2-dimensional $\kappa$-deformed space reflected by 
the algebra $\U_a$ with $\kappa=-\omega_0\,\log^{-1}(a)$ is in fact ``infinite dimensional" as a metric
noncommutative space. Theorem \ref{existence} is a tentative to restore a metric. For instance, the 
distances on the state space $\SS(\U_a)$ generated by Connes' formula
$$
d(\omega,\omega'):=\sup \set{\vert \omega(a)-\omega'(a)\vert \, :\, a\in \A ,\,  \vert \vert [\DD,a] \vert \vert \leq 1} 
,\qquad \omega,\omega' \in \SS(\U_a)
$$
are infinite in case $(ii)\,$of Theorem \ref{existence}, while in case $(iii)\,$some states can be at finite distances.
\end{remark}

\begin{remark}
In the noncommutative geometry given by a spectral triple $(\A,\H,\DD)$, the differentiability is defined by 
the operator $\DD$ in a subtle way:

In case $ii)$ of Theorem \ref{nonexistence}, let $\partial_\ell:=i[\DD_\ell,\cdot]$ be a closed unbounded 
$^*$-derivation from  $\B\big(\ell^2(G)\big)$ onto itself with $G=G_a$.  Then 
$\bigcap_{k\in \N} \text{Dom }(\partial_\ell^k)$ is a Fr\'echet sub-$^*$-algebra $\B_\ell$ of $\B(\ell^2(G))$ which 
is stable under smooth functional calculus and which contains $\C[G]$; thus 
$\mathcal{C}_\ell=\B_\ell \bigcap C^*(G)$ is a Fr\'echet sub-$^*$-algebra of $ C^*(G)=C_{red}^*(G)$ which is 
stable under smooth functional calculus and which contains $\C[G]$  \cite{CM1}. Since $G_a$ (being amenable 
with exponential growth) does not enjoy the Rapid Decay property (\cite[Corollary 3.1.8 ]{Jolissaint} and 
\cite[(1.3) Theorem]{Ji}), we can conclude that if $\H^\infty_\ell:=\bigcap_{k\in \N}\text{Dom }\DD_l^k$, then 
$\rho \,\big(\mathcal{C}_\ell \big) \neq \H^\infty_\ell$ where $\rho$ is the map: 
$x\in C_{red}^*(G) \mapsto x(\delta_e) \in \ell^2(G)$.
\end{remark}

\begin{remark}
The operator $\DD$ given in Theorem \ref{existence} $(iii)$ is not directly related to the group structure of 
$G_a$ but rather connected to the underlying dynamical system associated to the algebraic nature of $a$: it 
depends explicitly of the isomorphism-invariant $\set{c_q(a) \, : \,q\in \N^*}$.
\end{remark}

\section{Conclusion}

We have shown that $\kappa$-Minkowski space defined by \eqref{commutationkappa} can be reduced to a 
compact or discrete version. Depending on $\kappa$, or on $a$ defined in \eqref{defa}, this involves 
discrete amenable groups $G_a$, in particular the well-known Baumslag--Solitar ones. The associated 
$C^*$-algebras $\U_a$ can be viewed as a deformation of the two-torus. They are different when $a$ varies 
within the rational numbers (of zero Lebesgue measure) 
because of the structure of the underlying dynamical system. 
We also have shown that for quadratic algebraic $a$, the $K$-theory does not give the full classification of 
the algebras $\U_a$. 
For transcendental values of $a$, which are dense in $\R_+$ and of full Lebesgue measure, all these algebras 
are isomorphic to each other.

Due to the exponential growth of $G_a$, we have proved 
that the algebras $\U_a$ are not only quasidiagonal but also residually finite dimensional. 
They admit different spectral triples: the ones which are quasi-equivalent to the left regular representation 
and are never $p$-summable but only $\theta$-summable, \textsl{i.e.} they are of ``infinite metric dimension". 
This situation reminds us of the passage from non-relativistic to relativistic quantum mechanics: in quantum field 
theory, the $\theta$-summability (and not the $p$-summability) naturally occurs due to the behaviour of 
$\Tr(e^{-tH})$ (when $t\rightarrow 0$) where $H$ is the Hamiltonian (or $\DD^2$), see for instance \cite{CHKL}.

The other faithful representations can generate fancy spectral triples which can have arbitrary summability 
$p \geq 2$ (or ``dimension") depending on the algebraic properties of the real parameter $a$, 
but are in fact degenerate to some extent. Unfortunately, the topological content of these 
unbounded Fredholm modules may be trivial, in the sense that they correspond to trivial elements of $K$-homology 
(for instance, when $\lambda_q=1$, the associated bounded Fredholm module is degenerate). 

The nonexistence theorem, though powerful, does not preclude the possible existence of a genuine, 
non-degenerate, nontrivial spectral geometry on the $\kappa$-deformation spaces presented here, they only 
restrict the possible algebra representations that could be used in the construction.

This shows how delicate the notion of spectral or metric dimensions of  $\kappa$-Minkowski space is, and how 
subtle its analysis through noncommutative geometry.

\section*{Acknowledgments}

We thank Alain Connes, Michael Puschnigg, Adam Skalski and Shinji Yamashita for helpful discussions or 
correspondence.

B. I. and T. S. acknowledge the warm hospitality of the Institute of Physics at the Jagiellonian University in Krakow 
where this work was started under the Transfer of Knowledge Program ``Geometry in Mathematical Physics".

\end{document}